\documentclass[pre,twocolumn,superscriptaddress,nofootinbib]{revtex4}

\usepackage{graphicx,chemarr}

\begin{document}

\title{Statistics of correlated percolation in a bacterial community}

\author{Xiaoling Zhai}
\affiliation{Department of Physics and Astronomy, Purdue University, West Lafayette, Indiana 47907, USA}

\author{Joseph W.\ Larkin}
\affiliation{Division of Biological Sciences, University of California San Diego, Pacific Hall Room 2225B, Mail Code 0347, 9500 Gilman Drive, La Jolla, CA 92093, USA}

\author{Kaito Kikuchi}
\affiliation{Division of Biological Sciences, University of California San Diego, Pacific Hall Room 2225B, Mail Code 0347, 9500 Gilman Drive, La Jolla, CA 92093, USA}

\author{Samuel E.\ Redford}
\affiliation{Division of Biological Sciences, University of California San Diego, Pacific Hall Room 2225B, Mail Code 0347, 9500 Gilman Drive, La Jolla, CA 92093, USA}

\author{G\"urol M.\ S\"uel}
\affiliation{Division of Biological Sciences, University of California San Diego, Pacific Hall Room 2225B, Mail Code 0347, 9500 Gilman Drive, La Jolla, CA 92093, USA}
\affiliation{San Diego Center for Systems Biology, University of California San Diego, La Jolla, CA 92093, USA}

\author{Andrew Mugler}
\email{amugler@purdue.edu}
\affiliation{Department of Physics and Astronomy, Purdue University, West Lafayette, Indiana 47907, USA}

\begin{abstract}
Signal propagation over long distances is a ubiquitous feature of multicellular communities. In biofilms of the bacterium {\it Bacillus subtilis}, we recently discovered that some, but not all, cells participate in the propagation of an electrical signal, and the ones that do are organized in a way that is statistically consistent with percolation theory. However, two key assumptions of percolation theory are violated in this system. First, we find here that the probability for a cell to signal is not independent from other cells but instead is correlated with its nearby neighbors. We develop a mechanistic model, in which correlated signaling emerges from cell division, phenotypic inheritance, and cell displacement, that reproduces the experimental results. Second, we observed previously that the fraction of signaling cells is not constant but instead varies from biofilm to biofilm. We use our model to understand why percolation theory remains a valid description of the system despite these two violations of its assumptions. We find that the first violation does not significantly affect the spatial statistics, which we rationalize using a renormalization argument. We find that the second violation widens the range of signaling fractions around the percolation threshold at which one observes the characteristic power-law statistics of cluster sizes, consistent with our previous experimental results. We validate our model using a mutant biofilm whose signaling probability decays along the propagation direction. Our results identify key statistical features of a correlated percolating system and demonstrate their functional utility for a multicellular community.
\end{abstract}

\maketitle


\section{Introduction}
Long-range signal transmission is central to the function of many multicellular communities. However, cell-to-cell variability within these communities \cite{QB_Bochong, MC_Symmons} can cause some cells not to participate in signaling, which may degrade or attenuate the signal \cite{PTRS_Steinberg, NRN_Waxman, JosephCellSystem}. In physics, signal transmission in the presence of non-propagating agents is the domain of percolation theory \cite{stauffer2014introduction}. As a result, many investigators have turned to percolation theory to describe signal transmission in multicellular systems. In neuroscience, percolation theory has been used to describe (i) the transition from a fully connected to a disconnected electrical network in rat hippocampus cultures \cite{PRL_Breskin, PhyRep_Eckmann}, (ii) the spatiotemporal structure of viral propagation within astrocyte monolayers \cite{gonci2010viral}, and (iii) the transition from conscious to unconscious brain activities during general anesthesia \cite{zhou2015percolation}. In pancreatic islets, percolation theory has been used to understand the dependence of calcium wave propagation on the coupling strength of gap junctions between the islet cells \cite{BenningerBioPhysJ}. In colonies of {\it Spirostomum} (an aquatic worm-like cell), percolation theory was recently shown to describe how the propagation of a hydrodynamic cell-to-cell trigger-wave depends on the colony density \cite{mathijssen2018collective}.

We recently demonstrated that percolation theory governs the transmission of an electrical signal from the interior to the periphery of a biofilm of {\it Bacillus subtilis} bacteria \cite{JosephCellSystem}. In this system, starvation of the interior cells causes release of intracellular potassium, which leads to depolarization and potassium release in neighboring cells, resulting in a cell-to-cell relay wave that propagates to the biofilm periphery \cite{LiuNature, PrindleNature, Cell_Humphries}. The signal temporarily prevents peripheral cells from taking up nutrients and thus allows nutrients to diffuse to the interior cells, preserving biofilm viability and increasing its overall fitness \cite{LiuNature}. However, it turns out that not all cells participate in the potassium release: we discovered that the fraction of participating cells is near the percolation threshold, and that clusters of participating cells have a size distribution that follows a power law with an exponent predicted by percolation theory \cite{JosephCellSystem}. Operating near the percolation threshold allows the biofilm to maintain successful signal transmission while minimizing the number of cells that undergo the costly potassium release \cite{JosephCellSystem}.

\begin{figure*}
\centering
\includegraphics[width=\textwidth]{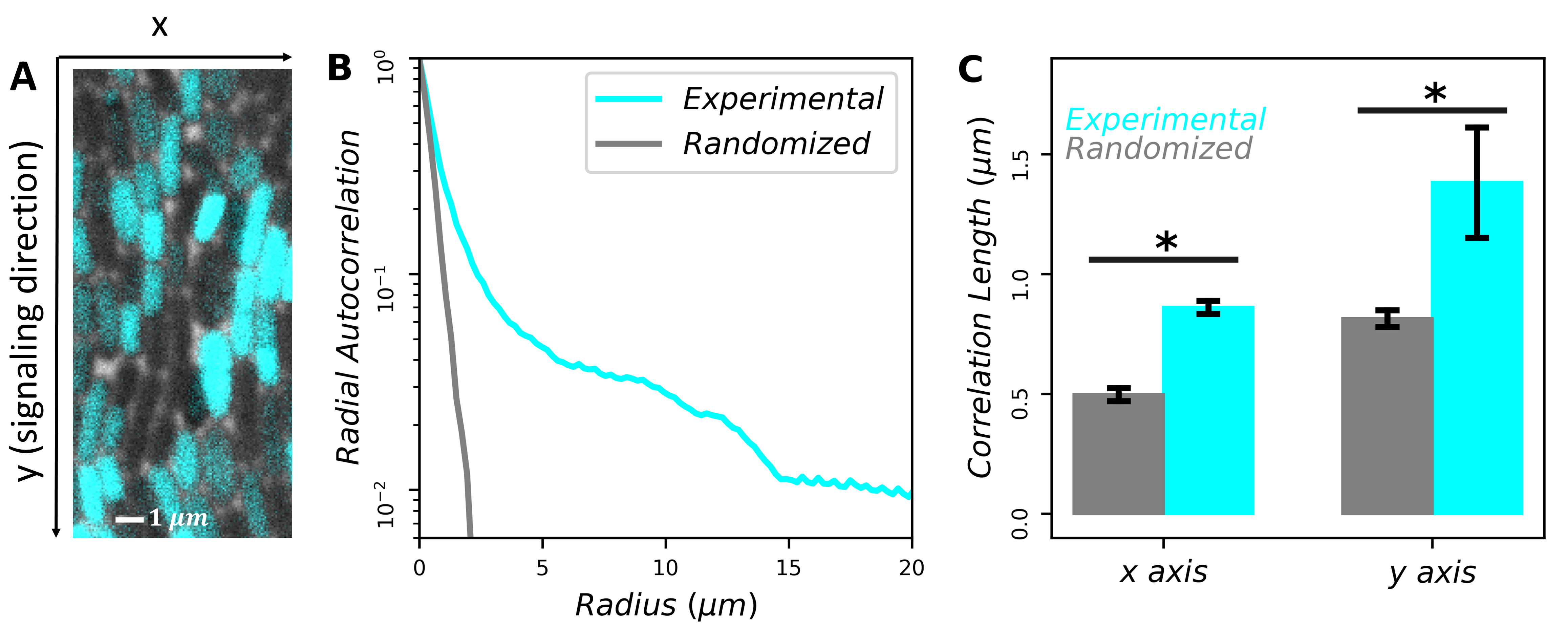}
\caption{Signaling probability of each cell is correlated with neighboring cells. (A) Zoomed-in snapshot of cells in biofilm during peak of signal transmission (actual experimental window is approximately 35 cells tall by 230 cells wide). Cyan indicates fluorescence intensity of ThT dye, proportional to membrane potential. (B) Correlation function is longer-range than that from randomized data ($N = 3$ biofilms). (C) Correlations are significantly longer than random both perpendicular ($x$) and parallel ($y$) to the signaling direction ($p < 0.001$ and $p = 0.007$ assuming Gaussian errors, respectively).}
\label{corr}
\end{figure*}

Despite the success of percolation theory as a description of signal transmission within this system, it is reasonable to suspect that several key assumptions of percolation theory may be violated. First, percolation theory assumes that the probability for each cell to participate in signal transmission is independent of other cells. However, in reality it may be that the participation probability of a cell is correlated with that of its neighbors. For example, if the molecular mechanism governing participation is heritable, then one expects the participation of a given cell to be correlated with other cells in its lineage, which are most likely to be nearby in the densely packed biofilm. Second, percolation theory assumes that the participation probability does not vary from one biofilm to another, or from location to location within a biofilm. However, in reality we know that there is variability across biofilms, and particular mutant strains have spatial variability in the participation fraction\cite{JosephCellSystem}. These considerations raise the question of when and how percolation theory remains a predictive description of signal transmission in biological systems, despite the fact that the assumptions of percolation theory may be violated by the system components.

Here we use a combination of simulations and experiments to investigate the statistical properties of signal percolation in a bacterial biofilm. We find that signal correlations exist between cells, due to a combination of phenotypic inheritance and spatial proximity of a cell to its progeny. We find that while these correlations lower the percolation threshold, they are not sufficiently long-range to affect the cluster size statistics. Instead, we find that variability in the signaling fraction across biofilms affects the statistics by widening the range of fractions at which one observes the power-law distribution of cluster sizes. We validate our findings using a mutant biofilm whose participation fraction decays as a function of propagation distance. Our results demonstrate that certain community-level signaling properties are robust to cell-level features whereas others are not, and we discuss the implications for biofilm function.

\section{Results}

\begin{figure*}
\centering
\includegraphics[width= .8\textwidth]{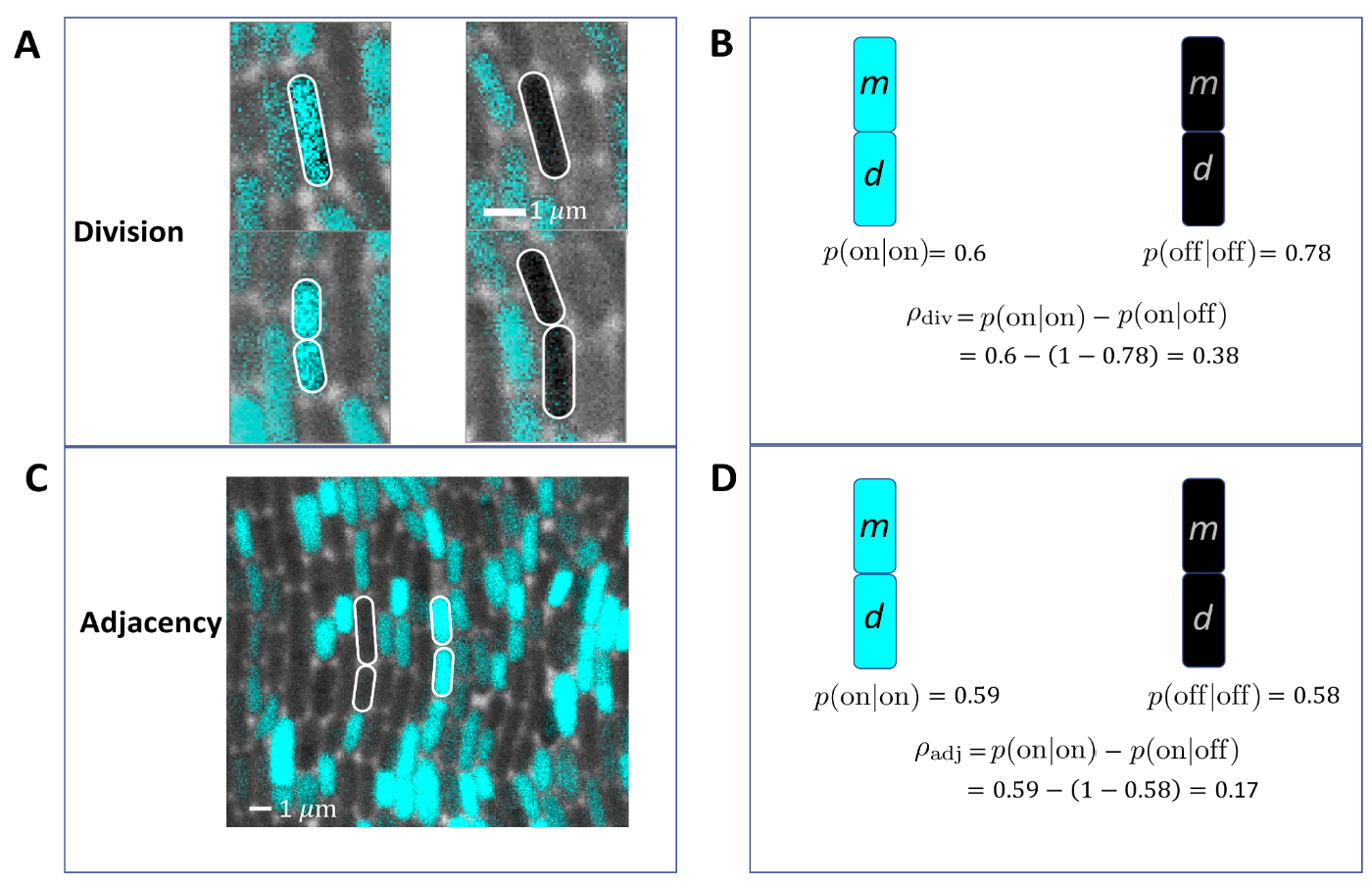}
\caption{Order parameter $\rho$ quantifies degree of spatial correlations. (A, B) Lineage-tracing experiments yield $\rho_{\rm lin} = 0.38$ ($N = 49$ division events). (C, D) Spatial analysis of the biofilm images yield $\rho_{\rm adj} = 0.17$ ($N = 51$ cell pairs).}
\label{rho}
\end{figure*}

We first review the key features of electrical signaling in the biofilm \cite{LiuNature, PrindleNature, Cell_Humphries, JosephCellSystem}, and those of percolation theory, as these features will motivate our present results. In our experiments, we measure the membrane potential of cells during the peak of signal transmission using a fluorescent dye (cyan in Fig \ref{corr}A; see Materials and Methods). We previously observed a bimodal distribution of dye intensity across cells \cite{JosephCellSystem}, which provides a threshold above or below which we define cells as ``on'' (participating in the signal) or ``off'' (not participating in the signal), respectively. This observation motivates our use of percolation theory, as percolation theory describes the connectivity and spatial statistics of systems on a lattice in which each cell has a probability $\phi$ to be on.

Our experiments focus on a 2D cell monolayer at the edge of the biofilm (see Materials and Methods). We previously found that cells are most likely to have six neighbors \cite{JosephCellSystem}. For an infinite 2D, six-neighbor lattice, percolation theory predicts that (i) a connected path of on-cells emerges above the critical value $\phi_c = 1/2$, and that (ii) at $\phi_c$, the distribution $P(n)$ of sizes $n$ of connected clusters of on-cells becomes a power law \cite{stauffer2014introduction}.

In the experiments, we image a finite window of approximately 35 by 230 cells (see Materials and Methods). Finite-size effects can change the value of $\phi_c$ at which connectivity sets in, which we call $\phi_c^{\rm conn}$ \cite{stauffer2014introduction}. Indeed, simulations predict that $\phi_c^{\rm conn} = 0.45$ in this finite geometry \cite{JosephCellSystem}. Finite size effects should not change the value of $\phi_c$ at which $P(n)$ becomes a power law, which we call $\phi_c^{\rm pow} = 1/2$, so long as $\sqrt{n}$ is sufficiently below the smaller lattice dimension. However, at larger $n$ values the distribution will deviate from a power law, even at $\phi_c^{\rm pow}$, due to finite-size effects.

We previously observed that the fraction of on-cells in the experiments is $\phi = 0.43 \pm 0.02$ (mean $\pm$ standard error), and that the distribution $P(n)$ of on-cell cluster sizes is a power law over three decades \cite{JosephCellSystem}. The fact that $\phi \approx \phi_c^{\rm conn}$ suggests that the system sits at the connectivity threshold. However, the fact that $\phi <\phi_c^{\rm pow}$ raises the question of why a power law is observed, particularly one with no apparent finite-size effects at large $n$. To address this question, as well as the broader question of what features of percolating systems are expected to be robust to the underlying assumptions about the components, we now investigate the effects of signal correlations and of variability in the signaling fraction.

\subsection{Spatial correlations in signal participation}

Percolation theory assumes that a fraction $\phi$ of on-cells are situated randomly in space. However, in the biofilm one might expect that on-cells are spatially co-located, for example if participating in the signal is a heritable phenotype. To determine whether there are spatial correlations in on-cells, we measure the radial autocorrelation function
\begin{equation}
\label{C}
C(r) = \langle s_i s_j\rangle_r - \phi^2,
\end{equation}
where $s = 1$ for on-cells, $s = 0$ for off-cells, and the average is taken over all pixels $i$ and $j$ whose separation is $r$ (see Materials and Methods). We find that $C(r)$ is a decreasing function of $r$, as expected (Fig \ref{corr}B, cyan curve). We then compare $C(r)$ to the autocorrelation function computed with the locations of on-cells randomized. Specifically, we retain the locations of all cells and the number of on-cells, but we randomize which cells are on (as would be the case in percolation theory). We see in Fig \ref{corr}B that $C(r)$ falls off more steeply in this case (gray curve). These results suggest that on-cells are more spatially correlated than expected from random placement.

We next investigate the strength of correlation perpendicular ($x$) and parallel ($y$) to the direction of signal transmission. We define the correlation lengths as $\xi_x = \int dx\ C(x)$ and $\xi_y = \int dy\ C(y)$, where $C(x)$ and $C(y)$ are defined as in Eq \ref{C} but restricted to horizontal ($x$) or vertical ($y$) separations, and the integrals run from zero to the maximal separation values. Even in the randomized data, we see that the correlation length is larger in the $y$ direction than in the $x$ direction (compare the gray bars in Fig \ref{corr}C) because cells are longer than they are wide, and the long axis of each cell is generally oriented in the signaling direction (Fig \ref{corr}A). In the actual (non-randomized) data, the correlation lengths are 70\% larger than random in both the $x$ and $y$ directions, and both differences are significant ($p < 0.01$; Fig \ref{corr}C). These results suggest that on-cells are significantly correlated both parallel and perpendicular to the signaling direction.

\begin{figure*}
\centering
\includegraphics[width= 1\textwidth]{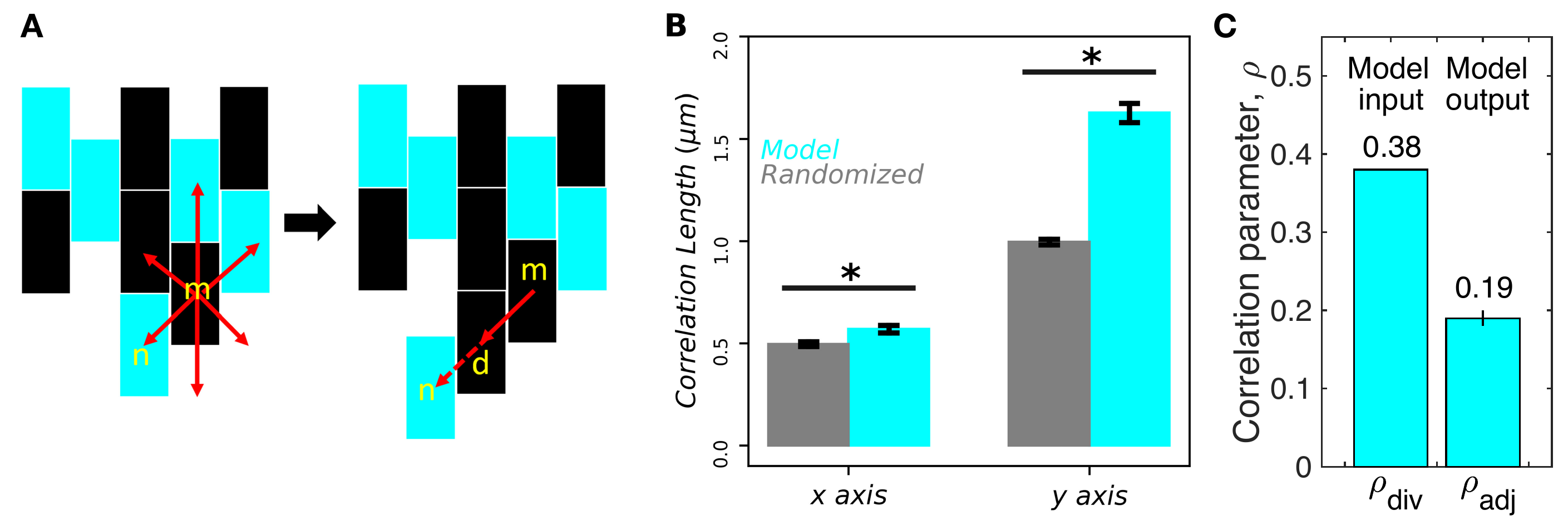}
\caption{Mechanistic model of correlated signaling captures experimental features. (A) Mother cell (m) produces daughter cell (d) with correlated signaling state at any neighboring site at which a maximum of one neighbor cell (n) is displaced. (B) Correlations are significantly longer than random both perpendicular ($x$) and parallel ($y$) to the signaling direction ($N = 10^3$ lattices; $p = 0.004$ and $p < 0.001$ assuming Gaussian errors, respectively). Compare to experiments in Fig \ref{corr}C. (C) Stochasticity in division times, neighbor selection, and cell displacement reduces correlation parameter from $\rho_{\rm div} = 0.38$ to $\rho_{\rm adj} = 0.19\pm 0.01$, close to experimentally measured $\rho_{\rm adj} = 0.17$ ($N = 10^3$ lattices).}
\label{model}
\end{figure*}

To quantify the correlation at the single-cell level, we consider the conditional probabilities $p({\rm on}|{\rm on})$ and $p({\rm off}|{\rm off})$, where $p({\rm on}|{\rm on})$ is the probability that a cell is on given that the cell directly upstream in the signaling direction is also on, and similarly for $p({\rm off}|{\rm off})$. We then calculate the order parameter
\begin{equation}
\label{rhodef}
\rho = p({\rm on}|{\rm on}) - p({\rm on}|{\rm off}),
\end{equation}
where $p({\rm on}|{\rm off}) = 1 - p({\rm off}|{\rm off})$. With no correlation, we have $p({\rm on}|{\rm on}) = p({\rm on}|{\rm off}) = \phi$, and therefore $\rho = 0$. With perfect correlation, we have $p({\rm on}|{\rm on}) = 1$ and $p({\rm on}|{\rm off}) = 0$, and therefore $\rho = 1$. Thus, $\rho$ quantifies the cell-to-cell correlation in the signaling direction on a scale from zero to one.

We estimate the conditional probabilities, and thus $\rho$, in two ways (Fig \ref{rho}). First, because cell division is usually parallel to the signaling direction, we track individual division events that occur prior to signaling (Fig \ref{rho}A; see Materials and Methods). We then count the number of times that the upstream daughter cell has the same signaling state as the downstream daughter cell. From this method we obtain $\rho_{\rm div} = 0.38$ (Fig \ref{rho}B). Second, we estimate the conditional probabilities directly from pairs of cells that are vertically adjacent during signaling (Fig \ref{rho}C; see Materials and Methods). From this method we obtain $\rho_{\rm adj} = 0.17$ (Fig \ref{rho}D). These results confirm at the single-cell level that spatial correlations exist in the signaling direction ($\rho_{\rm adj} > 0$) but suggest that these correlations are less strong than those produced directly by division ($\rho_{\rm adj} < \rho_{\rm div}$).

\subsection{Mechanistic model of correlated signaling}

To understand the experimental results above, we propose a mechanistic model of spatially correlated cell signaling. We hypothesize that the signaling state is heritable during cell division with a certain probability, and that cell displacement can occur at the leading edge as the biofilm grows. Specifically, as shown in Fig \ref{model}A, we generate a 2D, six-neighbor lattice of rectangular cells with aspect ratio 2 (the approximate experimental value) in the following way. Each cell divides after a time $\tau$ drawn from a Gaussian distribution with mean $\bar{\tau}$ and standard deviation $\delta\tau$. The ``mother'' cell (m) retains its location and signaling state, while the ``daughter'' cell (d) occupies one of the six neighboring locations with equal probability, given that the location is eligible. Eligibility requires that the neighboring location either be empty or be occupied by a neighboring cell (n) that, when displaced by the division along the same direction, would occupy an empty location (Fig \ref{model}A). The signaling state of the daughter, given that of the mother, is determined from the division parameter $\rho_{\rm div}$ and the fraction of on-cells $\phi$ according to
\begin{align}
\label{ponon}
p({\rm on}|{\rm on}) &= \phi + \rho_{\rm div} - \phi\rho_{\rm div}, \\
\label{ponoff}
p({\rm on}|{\rm off}) &= \phi - \phi\rho_{\rm div},
\end{align}
which follow from Eq \ref{rho} and the requirement that the fraction of on-cells remains $\phi$ throughout the process (see Materials and Methods). We produce a 100 by 230 lattice of cells by initializing the top row randomly and generating the next 99 rows according to the above mechanism. Then we remove the top 55 and bottom 10 rows, leaving a 35 by 230 cell window as in the experiments. This procedure allows the mechanism to achieve statistical steady state and focuses on the biofilm edge as in the experiments.

\begin{figure*}
\centering
\includegraphics[width= .75\textwidth]{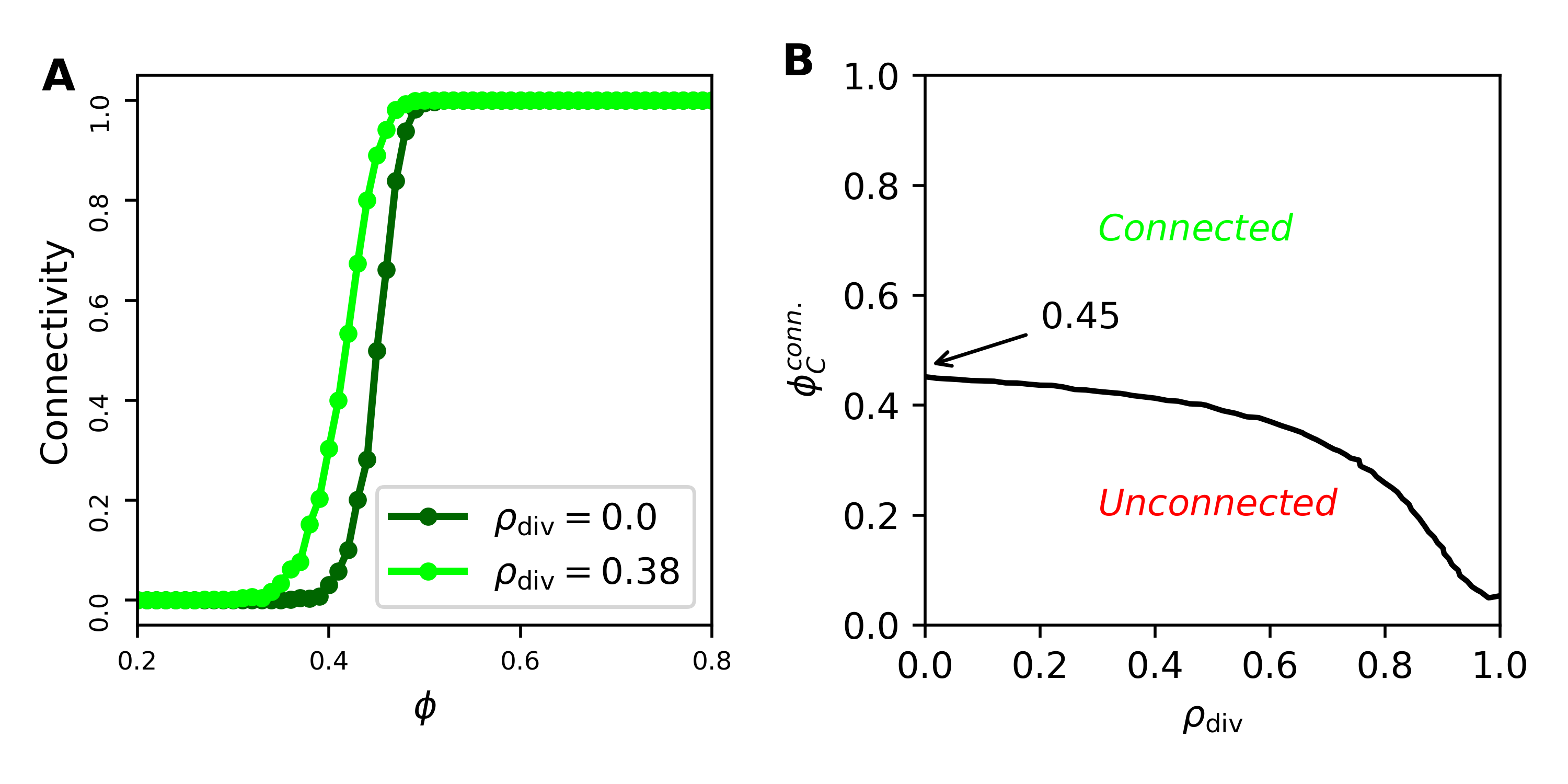}
\caption{Spatial correlations increase connectivity. (A) Connectivity, defined as probability that a connected path of on-cells exists, occurs at lower on-cell fraction $\phi$ as correlation parameter $\rho_{\rm div}$ increases ($N = 100$ lattices). (B) Connectivity threshold $\phi_c^{\rm conn}$, defined as $\phi$ value for which connectivity is $50\%$, decreases with $\rho_{\rm div}$ ($N = 100$ lattices).}
\label{conn}
\end{figure*}

We find that the spatial statistics are not sensitive to the value of $\delta\tau/\bar{\tau}$, so long as it is greater than zero, and therefore we average our results over the range $0 < \delta\tau/\bar{\tau} < 1$ (rejecting samples with $\tau \le 0$ for large $\delta\tau$). We also find that allowing neighbor cell displacement is necessary to generate correlations in the $x$ direction, but that allowing two or more levels of displacement does not qualitatively change the results. Thus, the only parameters in the model are $\phi$ and $\rho_{\rm div}$, which we set from the experiments as $\phi = 0.43$ \cite{JosephCellSystem} and $\rho_{\rm div} = 0.38$ (Fig \ref{rho}B).

This model, with no free parameters, makes five predictions. Specifically, as seen in Fig \ref{model}B, the model predicts that the correlation length in the $x$ direction is (i) significantly different from and (ii) 15\% higher than random, and that the correlation length in the $y$ direction is (iii) significantly different from and (iv) 70\% higher than random. Finally, as seen in Fig \ref{model}C, the model predicts that (v) the spatial correlation parameter measured from vertically adjacent cells after the biofilm is generated is $\rho_{\rm adj} = 0.19 \pm 0.01$, which is reduced from the model input value $\rho_{\rm div} = 0.38$ due to the stochasticity in division times, neighbor selection, and cell displacement.

Predictions (i) and (iii) are consistent with the experiments, as both the $x$ and $y$ correlation lengths were found to be significantly different than random (Fig \ref{corr}C). Prediction (iv), but not (ii), is consistent with the experiments, as both the $x$ and $y$ correlation lengths were found to be about 70\% higher than random (Fig \ref{corr}C). Finally, prediction (v) is consistent with the experiments, as $\rho_{\rm adj}$ was measured to be $0.17$ (Fig \ref{rho}D), which is very close to $0.19 \pm 0.01$. The fact that four out of five predictions are validated by the experiments gives us confidence that the model captures the basic underlying mechanism, especially because it has no free parameters.

\subsection{Impact of correlations on spatial statistics}

\begin{figure*}
\centering
\includegraphics[width= .75\textwidth]{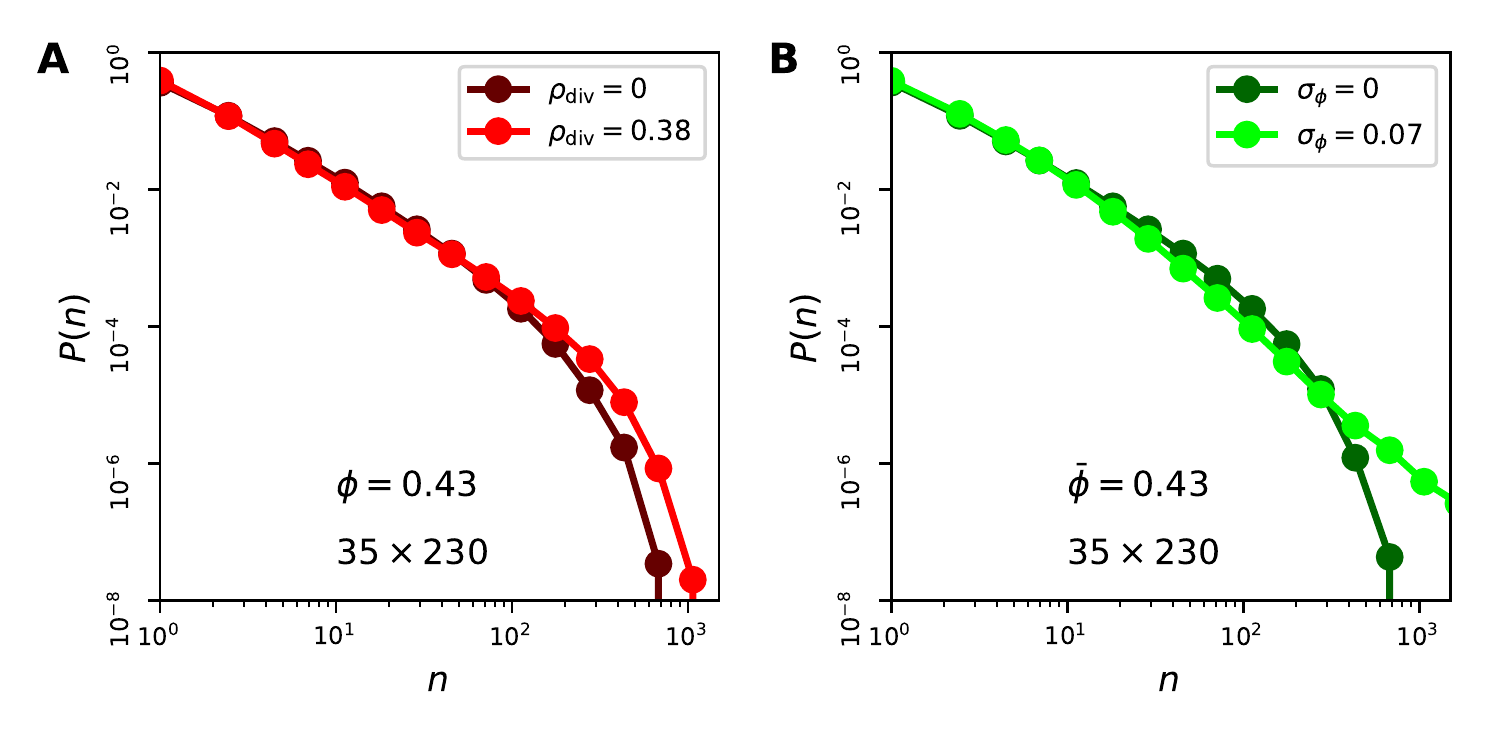}
\caption{Correlations cannot, but variability can, explain why $P(n)$ is a power law even though $\phi < \phi_c$. (A) Spatial correlations $\rho_{\rm div} = 0.38$ have little effect on distribution, in particular not removing exponential rolloff at large $n$ ($N = 100$ lattices). (B) Variability $\sigma_\phi = 0.07$ removes rolloff, causing distribution to approach a power law over three decades ($N = 10^3$ lattices).}
\label{pow}
\end{figure*}

We now use our mechanistic model to investigate the impact of the spatial correlations on the statistical properties of the biofilm. First we focus on the connectivity: the probability, over an ensemble of simulated biofilms, that a connected path of on-cells exists from the top to the bottom of the lattice. The connectivity is expected to show a sharp transition from $0$ to $1$ at a critical fraction of on-cells $\phi_c^{\rm conn}$. For an infinite lattice (in 2D with six neighbors), $\phi_c^{\rm conn} = 1/2$ \cite{stauffer2014introduction}. Finite size effects reduce the sharpness, but $\phi_c^{\rm conn}$ can still be defined as the value of $\phi$ for which the connectivity is $50\%$. For a finite lattice of the approximate size of the experimental window (35 cells tall by 230 cells wide), without correlations, we previously found $\phi_c^{\rm conn} = 0.45$ \cite{JosephCellSystem} (Fig \ref{conn}A, dark green curve). With correlations, using our mechanistic model with $\rho_{\rm div} = 0.38$, we find $\phi_c^{\rm conn} = 0.4$ (Fig \ref{conn}A, light green curve). More generally, the connectivity threshold is shown as a function of $\rho_{\rm div}$ in Fig \ref{conn}B, and we see that as $\rho_{\rm div}\to1$, $\phi_c^{\rm conn}$ becomes close to zero, even with the stochasticity inherent in the model. Thus, spatial correlations reduce the connectivity threshold. This makes sense, as correlations increase the probability of connected on-cells, particularly in the signaling direction, and this lowers the fraction of on-cells needed to created a connected path.

Second, we investigate the impact of correlations on the distribution $P(n)$ of sizes $n$ of connected clusters of on-cells. $P(n)$ is expected to become a power law at a critical fraction of on-cells $\phi_c^{\rm pow} = 1/2$ \cite{stauffer2014introduction}. The experimental fraction of on-cells is $\phi = 0.43 \pm 0.02$ \cite{JosephCellSystem}, which is lower than $\phi_c^{\rm pow}$. In simulations without correlations, at $\phi = 0.43$, we find that $P(n)$ acquires a rolloff (when viewed on a log-log scale) at large $n$ (Fig \ref{pow}A, dark red curve). The rolloff indicates that the distribution is becoming more exponential, as expected for $\phi < \phi_c^{\rm pow}$. However, in experiments, we find that $P(n)$ maintains the power law dependence, with no rolloff, for three decades, i.e.\ out to $n = 10^3$ \cite{JosephCellSystem}. Because we have seen that spatial correlations preserve connectivity at lower $\phi$ (Fig \ref{conn}), we hypothesize that correlations may also preserve the power law dependence of $P(n)$ at lower $\phi$, and thus explain the experimental observation. Surprisingly, using our mechanistic model, we find that the spatial correlations actually have little impact on $P(n)$ (Fig \ref{pow}A, light red curve): the rolloff is slightly shifted to larger $n$, but it is certainly still present over the three-decade range.

\begin{figure*}
\centering
\includegraphics[width= .75\textwidth]{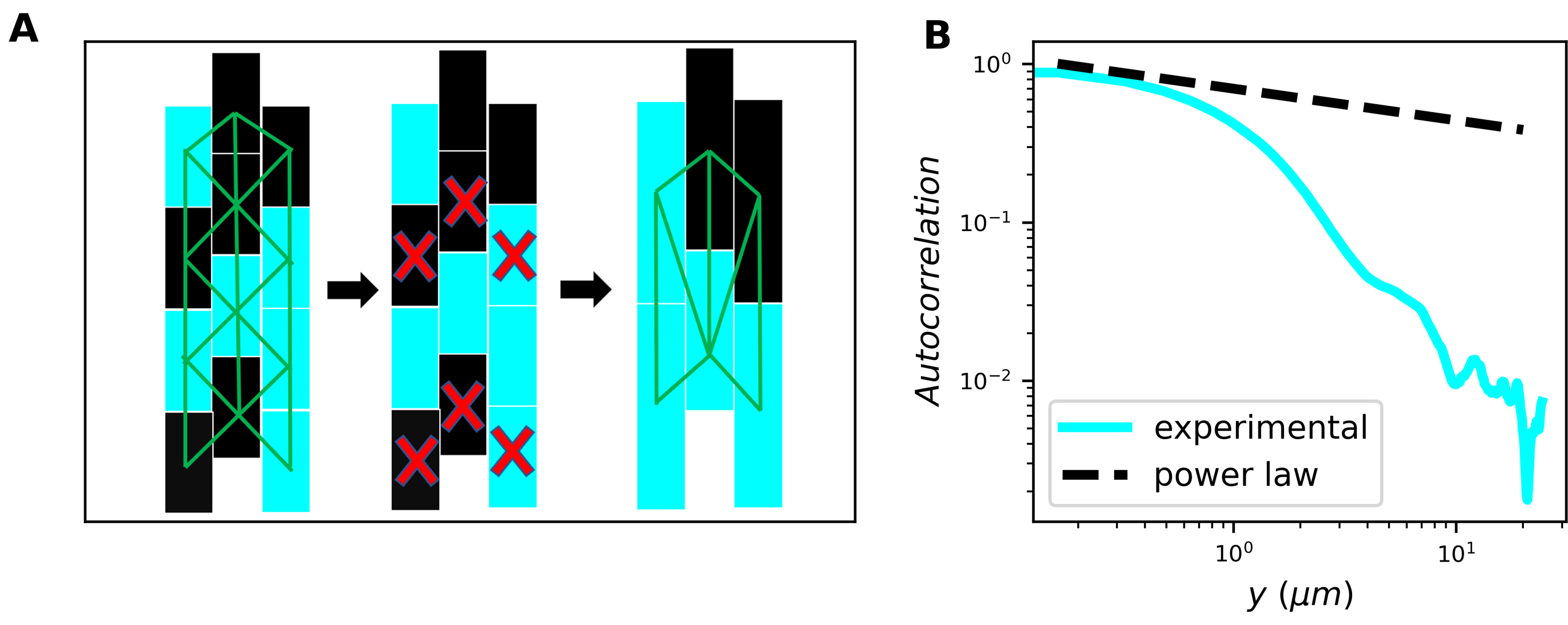}
\caption{Short-range correlations do not affect critical properties. (A) Illustration of the renormalization argument: upon site decimation, lattice remains triangular, $\phi$ remains constant, and $\rho$ vanishes. (B) Correlation function in experiments is short-range, i.e.\ sub-power-law ($N = 3$ biofilms).}
\label{renorm}
\end{figure*}

Why do correlations not change the distribution of cluster sizes? Renormalization-group arguments from statistical physics imply that correlations do not change the critical properties of percolation theory if the correlations are sufficiently short-range \cite{weinrib1984long}. The intuitive reason can be seen from a site-decimation procedure \cite{stauffer2014introduction}, as illustrated in Fig \ref{renorm}A. We imagine decimating every other cell in each column (red X's), with each remaining cell expanding to fill the space below it. Fig \ref{renorm}A illustrates that the resulting lattice remains triangular (green lines). Furthermore, because the probability of any cell to be on is $\phi$, the fraction of on-cells remains $\phi$ after decimation. Finally, the new conditional probabilities after one round of decimation are
\begin{align}
\label{p1def1}
p_1({\rm on}|{\rm on}) &= p({\rm on}|{\rm on})p({\rm on}|{\rm on}) + p({\rm on}|{\rm off})p({\rm off}|{\rm on}), \\
\label{p1def2}
p_1({\rm on}|{\rm off}) &= p({\rm on}|{\rm on})p({\rm on}|{\rm off}) + p({\rm on}|{\rm off})p({\rm off}|{\rm off}),
\end{align}
which follow from the rules of probability and the assumption that the signaling state is spatially Markovian, i.e.\ the daughter is conditionally independent of the grandmother given the mother (see Materials and Methods). As a result, the correlation parameter after one round of decimation is $\rho_1 = p_1({\rm on}|{\rm on}) - p_1({\rm on}|{\rm off}) = [p({\rm on}|{\rm on}) - p({\rm on}|{\rm off})]^2 = \rho^2$, where the first and last steps use the definition in Eq \ref{rhodef}, and the middle step inserts the expressions in Eqs \ref{p1def1} and \ref{p1def2} and simplifies (see Materials and Methods). Similarly, after $j$ rounds of decimation we have $\rho_j = \rho^{j+1}$. Because $\rho < 1$, we see that $\rho_j \to 0$ as $j\to\infty$. Thus, correlations vanish upon repeated rounds of decimation and renormalization. This means that correlations are not expected to change the critical properties of the distribution $P(n)$.

The above intuition only holds if the correlations are sufficiently short-range. Indeed, Eqs \ref{p1def1} and \ref{p1def2} assume that the correlations are minimally short-range, namely Markovian. In general, it has been shown that spatial correlations only affect the critical properties of percolation if they decay as a power law, specifically $C(r) \sim r^{-a}$ with $a > 3/2$ in 2D \cite{weinrib1984long}. As seen in Fig \ref{renorm}B, the correlations in the experimental data are much shorter-range than a power law. This suggests that the spatial correlations that we observe in the biofilm are not sufficiently long-range to affect the critical properties. Together with Fig \ref{pow}A, we conclude that spatial correlations are not sufficient to explain the experimentally observed power law dependence of $P(n)$ over three decades \cite{JosephCellSystem}.

\subsection{Variability in signaling fraction across biofilms}

If spatial correlations cannot explain the experimentally observed power law, then what can? An important feature of the experiments that is not yet accounted for in the model is the variability in the on-cell fraction $\phi$ across biofilms. In particular, we previously measured a standard deviation of $\sigma_{\phi}=0.07$ across 12 experiments (from which the standard error of $0.07/\sqrt{12} = 0.02$ comes) \cite{JosephCellSystem}. Therefore, using the model we investigate the effect of variability in $\phi$ across lattices on the distribution of cluster sizes $P(n)$. To do so, we draw $\phi$ for each lattice from a Gaussian distribution with standard deviation $\sigma_\phi$. Because we have found that correlations have little effect on $P(n)$, we set $\rho_{\rm div} = 0$ from here on for simplicity.

The results are shown in Fig \ref{pow}B, and we see that $\sigma_\phi$ has a significant effect on the distribution. In particular, for the experimental value $\sigma_{\phi}=0.07$ (light green curve), we see that the exponential rolloff at large $n$ is removed, extending the range of the power law out to $n \sim 10^3$ as observed in the experiments \cite{JosephCellSystem}. The intuitive reason is that a non-negligible fraction of lattices in the ensemble have $\phi$ values that are equal to or greater than $\phi_c^{\rm pow} = 1/2$. Because $\phi$ is higher in these lattices, they are more likely to have large clusters. Therefore, these lattices dominate the distribution at large $n$, eliminating the rolloff. Thus, variability in $\phi$ effectively widens the range of mean $\bar{\phi}$ values at which a power law distribution is observed. We conclude that the experimental variability in $\phi$ across biofilms is sufficient to explain the experimentally observed power-law distribution.

\subsection{Model validation using mutant strain}

\begin{figure*}
\centering
\includegraphics[width= .75\textwidth]{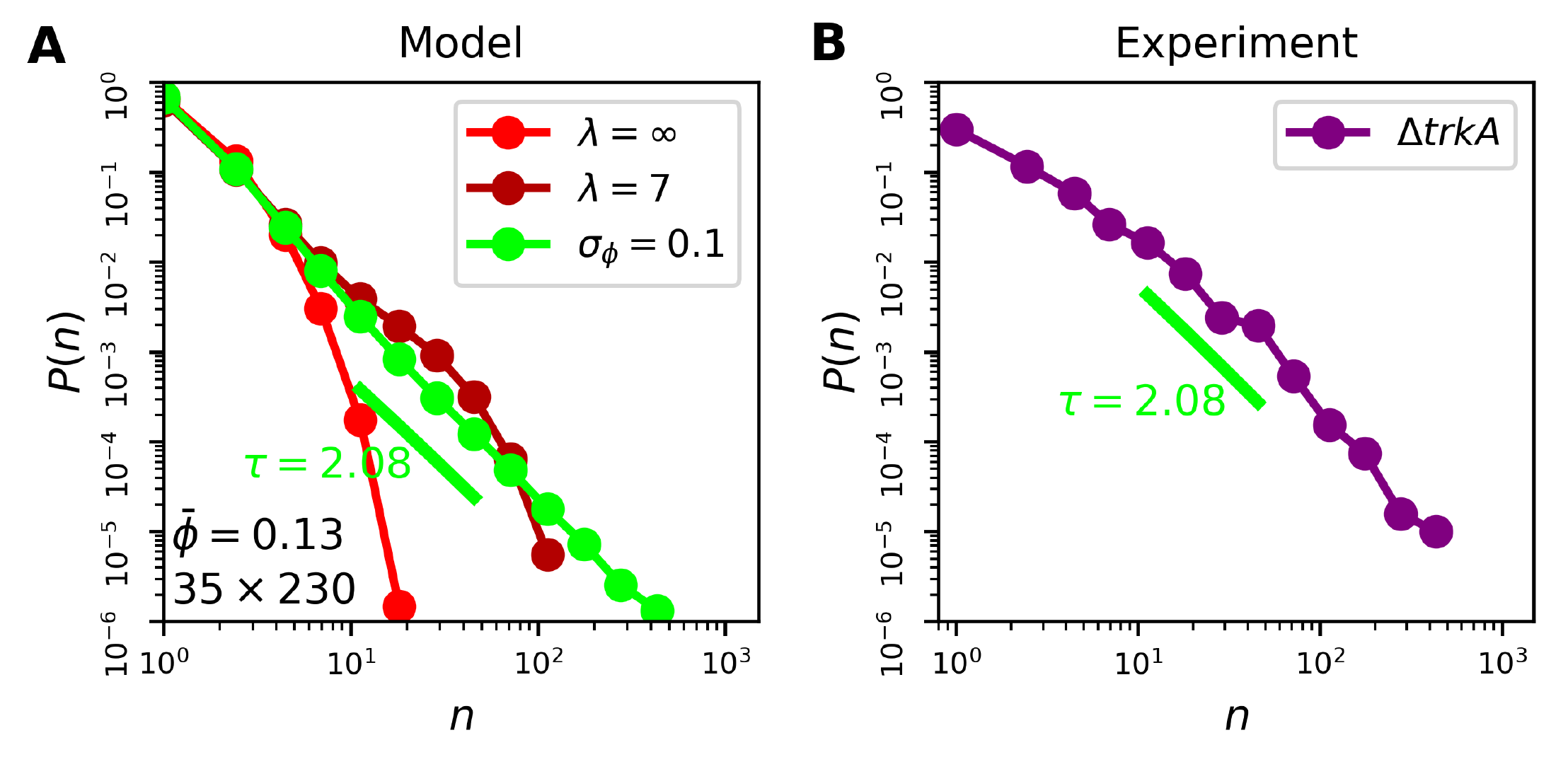}
\caption{Statistics of mutant $\Delta${\it trkA} strain. (A) We progressively incorporate into the model the on-cell fraction $\phi = 0.13$ (light red), the exponential decay of $\phi$ in space with lengthscale $\lambda = 7$ cells (dark red), and the variability $\sigma_\phi = 0.1$ across lattices (green); $N = 10^3$ lattices for each. Resulting $P(n)$ is a power law (green) despite the fact that $0.13$ is far below the critical fraction $\phi_c^{\rm pow} = 1/2$. (B) $P(n)$ from $\Delta${\it trkA} data is a power law whose exponent is consistent with the model ($N = 7$ biofilms).}
\label{trkA}
\end{figure*}

How can our model be tested with further experiments? One approach is to investigate a system with a different fraction of on-cells and see if our model remains valid. We previously investigated mutant strains with different on-cell fractions, including the $\Delta${\it trkA} strain with $\bar{\phi} = 0.13$ and $\sigma_\phi = 0.1$ \cite{JosephCellSystem}. As seen in Fig \ref{trkA}A (light red curve), basic percolation theory ($\rho_{\rm div} = 0$, $\sigma_\phi = 0$) predicts that a system with an on-cell fraction of $\phi = 0.13$ would have a distribution of cluster sizes $P(n)$ that is entirely exponential because $0.13$ is much lower than $\phi_c^{\rm pow} = 1/2$.

However, the $\Delta${\it trkA} strain differs from the wild-type strain in that the fraction of on-cells is not constant in space, but rather decreases along the signaling direction \cite{PrindleNature} with a characteristic lengthscale of approximately $\lambda = 15$ $\mu$m, or about $7$ cell lengths \cite{JosephCellSystem}. To incorporate this feature into the model, we allow the on-cell fraction to vary as $\phi(y)= \phi_0 e^{-y/\lambda}$, where $\phi_0$ is set to ensure that the spatial average of $\phi(y)$ is $0.13$. We see in Fig \ref{trkA}A (dark red curve) that this feature extends the distribution to larger cluster sizes $n$. The reason is similar to that given above regarding variability: the portions of the lattice in which $\phi$ is large contain large clusters, thereby enhancing the large-$n$ region of the distribution. Nonetheless, the distribution remains far from a power law in its shape. In particular, a clear exponential rolloff at large $n$ is evident.

If our main finding above is correct, namely that variability in $\phi$ across biofilms is a crucial determinant of the shape of $P(n)$, then we must also incorporate into our model the variability $\sigma_\phi = 0.1$ observed for the $\Delta${\it trkA} strain. Indeed, we find that doing so has a major effect on the distribution (Fig \ref{trkA}A, green curve). Specifically, it removes the exponential rolloff, resulting in a power-law distribution over almost three decades. This is a strong prediction, considering that $\bar{\phi} = 0.13$ is much lower than $\phi_c^{\rm pow} = 1/2$, and that without variability the shape is far from a power law even after accounting for the spatial dependence of $\phi$.

To test this prediction, we measure the distribution of cluster sizes in the $\Delta${\it trkA} biofilms (see Materials and Methods). Remarkably, the result, shown in Fig \ref{trkA}B, is a distribution that is roughly a power law over almost three decades, consistent with the model prediction. Indeed, the power-law exponent of $2.08$ estimated from the model distribution via a maximum likelihood technique \cite{clauset2009power} (Fig \ref{trkA}A, green line) is consistent with the slope of the experimental distribution (Fig \ref{trkA}B, green line). This result validates our model. In particular, it supports the finding that variability of the signaling fraction across biofilms plays an important role in shaping the statistical properties of the system.

\section{Discussion}
We have shown that experimentally observed features that go beyond the basic assumptions of percolation theory, including spatial correlations, variability, and non-uniformity, can have important consequences for signal propagation in a bacterial community. Using a mechanistic model that accounts for heritability in a cell's propensity to participate in signaling, we have found that signal correlations decrease the fraction of participating cells needed to create a connected path, but have little effect on the cluster statistics. In contrast, variability of the signaling fraction across samples has a significant effect on the statistics, in particular producing a power-law distribution of cluster sizes at signaling fractions lower than the expected critical fraction from percolation theory. We have validated our model using a mutant strain, in particular finding that both spatial decay and variability in the signaling fraction play a crucial role in shaping the signaling statistics.

We found that incorporating the experimentally observed variability and non-uniformity of the signaling fraction into the model was necessary to explain the experimentally observed cluster statistics, whereas incorporating the experimentally observed spatial correlations in signaling was not necessary. This finding implies that certain underlying cell-level features are important in determining population-level statistical properties, whereas others are not. This categorization is consistent with approaches from statistical physics, particularly the renormalization group, which reflect the powerful notion that some microscopic details are relevant for macroscopic properties, whereas others are provably irrelevant \cite{goldenfeld2018lectures}. Indeed, we explain our finding that spatial correlations do not affect the cluster statistics using a renormalization argument (Fig \ref{renorm}), as well as more rigorous known results from statistical physics \cite{weinrib1984long}. It will be interesting to see what other cell-level features are relevant or irrelevant for capturing population-level phenomena in multicellular systems.

Finite-size effects play an important role in our results. In particular, our experimental observation window is sufficiently short in the signaling direction ($\sim$35 cells) that spatial correlations in the signaling propensity have a measurable effect on the connectivity (Fig \ref{conn}). Yet, the window is wide perpendicular to signaling ($\sim$230 cells), and thus the window area is sufficiently  large that the spatial correlations have little effect on the cluster size statistics. This choice of window size follows from experimental constraints and the desire to focus on the short and wide biofilm edge, where signaling is most important for function \cite{PrindleNature}. Nonetheless, it is an interesting open question how the finite size and aspect ratio of the system set distinct thresholds for the relevance of correlations to the connectivity and cluster statistics.

Dimensionality also plays an important role in our results. Because the biofilm edge is where cell growth is most pronounced, it is quasi-two-dimensional. Therefore our experiments have focused on 2D monolayers of cells. However, the properties of percolation theory depend critically on the dimensionality of the system \cite{stauffer2014introduction}. In particular, the percolation threshold is generally smaller in 3D lattices than in 2D lattices \cite{wang2013bond} because there are more available paths for the signal to take. This observation suggests that a lower fraction of signaling cells is necessary in the bulk of the biofilm than at its edge. This prediction is currently difficult to test, as the 2D nature of our experiments is crucial for obtaining fluorescence data at the single-cell level.

The fact that spatial correlations lower the connectivity threshold in a finite system may help explain why the biofilm has an on-cell fraction of $\phi= 0.43 \pm 0.02$ \cite{JosephCellSystem}. Naive percolation theory predicts a threshold of $\phi_c^{\rm conn} = 1/2$ \cite{stauffer2014introduction}, which the biofilm does not meet. Accounting for finite-size effects lowers the threshold to $\phi_c^{\rm conn} = 0.45$ \cite{JosephCellSystem} (Fig \ref{conn}), which the biofilm barely meets. Accounting for correlations lowers the threshold further to $\phi_c^{\rm conn} = 0.4$ (Fig \ref{conn}), which the biofilm meets comfortably. Thus, correlations provide some leeway between the necessary and observed signaling fraction, which may enhance the reliability of signaling or make it robust to errors.

Finally, our study motivates further avenues of exploration in both statistical physics and cell biology. In statistical physics, our study motivates more general investigations of whether and how particular microscopic features affect macroscopic properties of percolation. The effects of spatial correlations in the site occupation probability are relatively well understood \cite{prakash1992structural, sahimi1996scaling, makse1998modeling, sahimi1998non}, whereas the effects of variability and non-uniformity in the site occupation probability are still relatively open questions \cite{kundu2016percolation, ikeda1979percolation}. In cell biology, our study builds on previous work \cite{PRL_Breskin, PhyRep_Eckmann, gonci2010viral, zhou2015percolation, BenningerBioPhysJ, mathijssen2018collective, JosephCellSystem} that demonstrates the utility of percolation theory as a quantitative and predictive description of multicellular phenomena. It will be interesting to see in what biological systems ideas from percolation theory will provide useful insights next.

\section{Materials and methods}
\subsection{Experimental methods}
\subsubsection{Microfluidics and experimental conditions}
Bacterial strains and growth conditions were as in \cite{JosephCellSystem}. We performed experiments in Y04D microfluidic plates using the CellASIC ONIX microfluidic system (EMD Millipore). Cells were imaged at the edge of biofilms and were confined to a single-cell layer by the PDMS structures of the microfluidic chamber. Each microscope field of view was roughly 330 $\mu$m $\times$ 70 $\mu$m and contained 8,000$-$10,000 cells. Every 5 minutes, we took phase contrast and fluorescence images on an Olympus IX83 inverted microscope with autofocus and a 40X, 0.6 NA air objective.

To probe membrane potential, we used the cationic fluorescent dye Thioflavin-T (ThT), which acts as a Nernstian voltage indicator \cite{PrindleNature}. When cells are hyperpolarized, they retain more of the dye and have a higher signal. ThT was present in the media at a concentration of 10 $\mu$M. We considered a cell to be an on-cell if its mean ThT signal exceeded a particular threshold during a signal pulse \cite{JosephCellSystem}.

\subsubsection{Computation of correlation function}
To compute correlation functions, we first thresholded ThT images so that they were binary: biofilm regions above the ThT threshold would appear white and sub-threshold regions would appear black. We then applied a 2-pixel radius median filter to thresholded images so that clusters of on-cells became contiguous white regions. From this image, we created a 2D autocorrelation plot using the ImageJ command FD Math. The resulting plot was mean-subtracted and normalized such that the origin had a value of 1 and decayed to 0 away from the origin (see source code for the Radially Averaged Autocorrelation ImageJ plugin for further details).

To compute the radial autocorrelation curves (Fig \ref{corr}B), we took a radial average of this 2D correlation plot. For $x$ and $y$ correlation curves, we took profiles of the correlation plot along the $x$ and $y$ axes, respectively.

To construct randomized images for such correlation computations, we took segmented biofilm images and randomly assigned a fraction of cells to be on and made them white. We then computed the autocorrelation curve on these images the same way as with the experimental images.

\subsubsection{Lineage tracing for $\rho_{\rm div}$}
To determine $\rho_{\rm div}$, we tracked individual cell lineages over time within biofilms using the mTrackJ imageJ plugin \cite{meijering2012methods}. For each lineage, we recorded the firing state (i.e.\ on or off) of the parent cell and the daughter cells. Using many lineages, we computed the conditional probabilities $p({\rm on}|{\rm on})$, $p({\rm on}|{\rm off})$, $p({\rm off}|{\rm on})$, and $p({\rm off}|{\rm off})$. We then computed the order parameter $\rho_{\rm div}$ using Eq \ref{rhodef}.

\subsubsection{Spatial analysis for $\rho_{\rm adj}$}
To determine $\rho_{\rm adj}$, we segmented cells in static images taken during signal pulses and determined the firing state of each cell (i.e.\ on or off). Because the electrical signal propagates in the direction of cell growth, cells are generally oriented along the signaling direction (Fig \ref{corr}A). We considered the upstream end of each cell to be the top and the downstream end to be the bottom. The adjacent cell in each case was defined as the cell whose bottom edge was closest to the given cell's top edge, and whose centroid was within half the average cell width. We then computed the conditional probabilities $p({\rm on}|{\rm on})$, $p({\rm on}|{\rm off})$, $p({\rm off}|{\rm on})$, and $p({\rm off}|{\rm off})$ for the firing state of a cell given the state of the adjacent cell. We then computed the order parameter $\rho_{\rm adj}$ using Eq \ref{rhodef}.

\subsubsection{Image analysis for $\Delta${\it trkA}}
We evaluated the cluster size distribution for $\Delta${\it trkA} biofilms in Fig \ref{trkA}B by first segmenting single biofilm cells in phase images using the Trainable Weka Segmentation plugin in ImageJ. We then thresholded the corresponding ThT images as described in the above section on computing correlation curves. Each contiguous white region in the thresholded image was a cluster of on-cells. We then counted how many segmented cells had the majority of their area within each cluster. The curve in Fig \ref{trkA}B plots the normalized histogram of these cluster sizes.

\subsection{Theoretical methods}
\subsubsection{Mechanistic model}
To derive Eqs \ref{ponon} and \ref{ponoff}, we require that the fraction of on-cells is $\phi$ at each step in the growth process. Specifically, the rules of probability state that
\begin{equation}
\label{prob1}
p(d) = \sum_m p(d,m) = \sum_m p(d|m)p(m),
\end{equation}
where $d$ is the signaling state (on, off) of the daughter, and $m$ is the signaling state (on, off) of the mother. Taking $d = $ on and requiring that $p({\rm on}) = \phi$ and $p({\rm off}) = 1-\phi$, Eq \ref{prob1} becomes
\begin{equation}
\phi = p({\rm on}|{\rm on})\phi + p({\rm on}|{\rm off})(1-\phi).
\end{equation}
Solving for $\phi$, we obtain
\begin{equation}
\phi = \frac{p({\rm on}|{\rm off})}{1+p({\rm on}|{\rm off})-p({\rm on}|{\rm on})}.
\end{equation}
Combining this equation with Eq \ref{rho} and solving for the conditional probabilities, we obtain Eqs \ref{ponon} and \ref{ponoff}.

\subsubsection{Renormalization argument}
To derive Eqs \ref{p1def1} and \ref{p1def2}, we recognize that the conditional probability of the daughter given the mother after one round of decimation is the conditional probability of daughter given the grandmother before the decimation. Again using the rules of probability, we write the latter as
\begin{equation}
\label{prob2}
p(d|g) = \sum_m p(d,m|g) = \sum_m p(d|m,g) p(m|g),
\end{equation}
where $g$ is the signaling state (on, off) of the grandmother. The spatial Markovian assumption states that $d$ is conditionally independent of $g$ given $m$. Therefore we have $p(d|m,g) = p(d|m)$, and Eq \ref{prob2} becomes
\begin{equation}
p(d|g) = \sum_m p(d|m) p(m|g).
\end{equation}
Setting $d = $ on and $g = $ on gives Eq \ref{p1def1}. Setting $d = $ on and $g = $ off gives Eq \ref{p1def2}.

To derive the relation $\rho_1 = \rho^2$ below Eq \ref{p1def2}, we insert Eqs \ref{p1def1} and \ref{p1def2} into the definition $\rho_1 = p_1({\rm on}|{\rm on}) - p_1({\rm on}|{\rm off})$. Using the shorthand $q \equiv p({\rm on}|{\rm on})$ and $r \equiv p({\rm on}|{\rm off})$, and recognizing that $p({\rm off}|{\rm on}) = 1-q$ and $p({\rm off}|{\rm off}) = 1-r$, this insertion obtains
\begin{equation}
\rho_1 = q^2 + r(1-q) - qr - r(1-r) = q^2 - 2qr + r^2 = (q-r)^2.
\end{equation}
Because $\rho = q-r$ (Eq \ref{rhodef}), we see that $\rho_1 = \rho^2$.
\\

\acknowledgments
We thank Aleksandra Walczak and Jordi Garcia-Ojalvo for helpful discussions.
This work was supported by the National Institute of General Medical Sciences (R01 GM121888 to G.M.S.\ and A.M.) and the Simons Foundation (376198 to A.M.).


\end{document}